\documentclass[fleqn,usenatbib]{mnras}

\usepackage[utf8]{inputenc}

\usepackage{acronym}
\usepackage{ae,aecompl}
\usepackage{booktabs}
\usepackage{dcolumn}
\usepackage{graphicx}
\usepackage{amsmath,amsfonts,amssymb}
\usepackage{mathrsfs}
\usepackage[normalem]{ulem}
\usepackage{epstopdf}

\usepackage[T1]{fontenc}

\let\oldhref\href
\renewcommand{\href}[2]{\oldhref{#1}{\hbox{#2}}}

\title[Retrieval of spectra for CCSN neutrinos]{Retrieval of energy spectra for all flavor of neutrinos from core-collapse supernova with multiple detectors}

\author[H. Nagakura et al.]{
Hiroki Nagakura$^{1}$\thanks{E-mail: hirokin@astro.princeton.edu}
\\
$^{1}$Department of Astrophysical Sciences, Princeton University, 4 Ivy Lane, Princeton, NJ 08544, USA
}

\date{Accepted XXX. Received YYY; in original form ZZZ}

\pubyear{2020}

\begin{document}
\label{firstpage}
\pagerange{\pageref{firstpage}--\pageref{lastpage}}
\maketitle

\begin{abstract}
We present a new method by which to retrieve energy spectrum for all flavor of neutrinos from core-collapse supernova (CCSN). In the retrieval process, we do not assume any analytic formulae to express the energy spectrum of neutrinos but rather take a direct way of spectrum reconstruction from the observed data; the Singular Value Decomposition algorithm with a newly developed adaptive energy-gridding technique is adopted. We employ three independent reaction channels having different flavor sensitivity to neutrinos. Two reaction channels, inverse beta decay on proton and elastic scattering on electrons, from a water Cherenkov detector such as Super-Kamiokande (SK) and Hyper-Kamiokande (HK), and a charged current reaction channel with Argon from the Deep Underground Neutrino Experiment (DUNE) are adopted. Given neutrino oscillation models, we iteratively search the neutrino energy spectra at the CCSN source until they provide the consistent event counts in the three reaction channels. We test the capability of our method by demonstrating the spectrum retrieval to a theoretical neutrino data computed by our recent three-dimensional CCSN simulation. Although the energy spectrum with either electron-type or electron-type anti-neutrinos at the CCSN source has relatively large error compared to that of other species, the joint analysis with HK + DUNE or SK + DUNE will provide precise energy spectrum of all flavors of neutrinos at the source. Finally, we discuss perspectives for improvements of our method by using neutrino data of other detectors.
\end{abstract}


\section{Introduction}\label{sec:intro} 
Kamiokande \citep{1987PhRvL..58.1490H} and IMB \citep{1987PhRvL..58.1494B} detected $\sim 20$ neutrinos of a few tens of MeV from SN 1987A, which is the first and only core-collapse supernova (CCSN) whose burst of neutrinos was directly observed. An important information on the CCSN environment can be extracted from the signal; the total energy of emitted neutrinos is the order of $10^{53} {\rm erg}$, which is consistent with our basic picture of CCSN. The detected neutrinos, however, seem to be only electron-type anti-neutrinos ($\bar{\nu}_e$), indicating that the energy of other flavors were less constrained. The statistics is also paltry, which prevents us from ascertaining the energy spectrum and its time evolution (but see \citet{2002PhRvD..65f3002L,2005PhRvD..72f3001M,2007PhRvD..76h3007Y,2015JPhG...42a3001V}), implying that the detailed characteristics of the neutrino signals remains largely unconstrained.

Since 1987, the scale and sensitivity of neutrino detectors have been improved considerably (see \citet{2018JPhG...45d3002H} and references therein for a recent review). Super-Kamiokande (SK), which is a current operating water Cherenkov detector, is capable of detecting $\sim 10,000$ neutrinos for CCSNe at a distance of $10$ kpc \citep{2007ApJ...669..519I}. The flavor dependent information may be obtained through utilizing its multiple reaction channels: inverse beta decay on proton (IBD-p), charged current reactions on Oxygen, elastic scattering on electrons (eES) and that on Oxygen \citep{2008JCAP...12..006M,2017JCAP...11..036G}, although the reaction channels except for IBD-p are feeble interactions for the typical energy of CCSN neutrinos, implying that the neutrino signals extracted from those channels will be low statistics. On the other hand, the issue will be alleviated in the future. For instance, \citet{2018JCAP...04..040G} suggested that spectrum reconstructions on mu- and tau- neutrinos (hereafter, they are denoted as $\nu_{\mu}$ and $\nu_{\tau}$, respectively, and being sometimes collectively expressed as $\nu_{x}$) are possible through a statistical approach by using the multiple reaction channels in a future water Cherenkov detector, Hyper-Kamiokande (HK). The project has been officially approved very recently and it will be fully operational from 2027.

A future liquid scintillator detector, JUNO \citep{2016JPhG...43c0401A}, which is currently under construction and planned to take data from 2021, will be also useful to probe each flavor information. They are the most sensitive to $\bar{\nu}_e$ via IBD, and the channel of neutrino proton elastic scattering (pES) may bring us to measure the energy spectrum of $\nu_x$ \citep{2002PhRvD..66c3001B,2011PhRvD..83k3006D}. Recently, \citet{2019PhRvD..99l3009L} demonstrated that energy spectra for all flavors of CCSN neutrinos can be retrieved by utilizing its three reaction channels: IBD-p, pES and eES. The obtained $\nu_e$ and $\nu_x$ spectra are, however, rather noisy due to low statistics even in the case with very nearby CCSN ($<1$ kpc), and they also suffer from a large systematic bias of energy threshold on the pES channel in the energy range of $\lesssim 20$MeV, which corresponds to the most important energy range for CCSN neutrinos, though \citep[see also][]{2011PhRvD..83k3006D}.

The retrieval of energy spectrum of all flavors of neutrinos can be done more accurately if coincident neutrino data from multiple detectors with different flavor sensitivities are available\footnote{We refer readers to \citet{2012ARNPS..62...81S} and references therein for the facilities having capabilities to contribute analyses of neutrinos from CCSNe. We also refer readers to previous works \citep[see, e.g.,][]{2018PhRvD..97b3019N} to retrieve energy spectrum of all flavors of neutrinos by using data of multiple detectors.}. In SK, HK and JUNO, for instance, they are sensitive to $\bar{\nu}_e$ through IBD reactions, meanwhile the Deep Underground Neutrino Experiment, DUNE, which is a proposed neutrino experiment with $\sim 40$ kton liquid-argon time-projection-chamber \citep{2016arXiv160105471A,2016arXiv160807853A,2020arXiv200806647A}, will provide high statistics data of electron-type neutrinos ($\nu_e$) via charged current reaction with Argon (CCAre). It is also planned to be operational in 2027, the same year as that of HK; hence, the simultaneous observations of CCSN neutrinos by these large-scale detectors may be realized in near future. It is, hence, interesting to consider how we can maximize the scientific return by combining the data of multiple detectors, which is the main subject in this paper.

Given theoretical models of CCSN and detector configurations, we can simulate data analyses of neutrino signals from CCSNe, and then study what information on CCSN dynamics can be extracted from them. One of the theoretical models of the neutrino signal frequently used in the literature is the so called "Garching model" \citep{2010PhRvL.104y1101H}, in which they numerically simulated an electron-capture $8.8 M_{\sun}$ supernova under spherically symmetric geometry; its time-dependent neutrino signals are publicly available. \citet{2013ApJS..205....2N} also provided theoretical templates of neutrino signals from CCSNe with various mass progenitors based on their spherically symmetric CCSN simulations with full Boltzmann neutrino transport. Employing the Nakazato models, \citet{2019ApJ...881..139S} estimated their neutrino count rates at SK with covering the time duration of $\sim 10$ s. More recently, the characteristics of late time neutrino signal ($> 10$ s) starts to be investigated based on spherically symmetric CCSN models numerically \citep{2020arXiv200804340W} and analytically \citep{2020arXiv200807070S}.

 Similar analyses are now possible to any numerical CCSN models by using a detector simulation software, SNOwGLoBES\footnote{SNOwGLoBES is available at \url{https://webhome.phy.duke.edu/~schol/snowglobes/} .}, in which detector responses to many reaction channels are provided \citep[see e.g.,][for more details.]{2012ARNPS..62...81S,2018JPhG...45a4002S}. For instances, \citet{2013ApJ...762..126O} employed SNOwGLoBES to estimate IBD-p count rate in a water Cherenkov detector like SK to their spherically symmetric CCSN models: \citet{2018MNRAS.480.4710S} developed an analysis pipeline based on SNOwGLoBES and estimated neutrino count rate on multiple detectors to their spherically symmetric and axisymmetric CCSN models: \citet{2019arXiv191203328W} carried out a systematic study of neutrino signals for 600 CCSN numerical models by SNOwGLoBES and analyzed their correlation to gravitational waves: very recently we studied the detailed neutrino signals based on our three-dimensional (3D) CCSN models with SNOwGLoBES \citep{2020arXiv200705000N}. As such, the capability of these detector software has been well matured and it still keeps evolving. Those efforts will further fill the gap between theoretical models to the observations.

Although those detector simulations are very useful, the actual data analysis is more complicated. One of the practical goals in the data analysis is to retrieve energy spectra of neutrinos from observed data, which will be performed by taking unfolding or statistical processes to the data of each reaction channel. The neutrino signal is, however, inevitably smeared out by various effects; for instances, detector responses may be one of the primary causes of the smearing, which depends on the instrument and reaction channel. This issue can be handled with a so-called response matrix to the injected signals, which is usually determined through Monte Carlo simulations of detector responses. It should be noted, however, that the unfolding process is not easy, since the spectrum inversion by using the response matrix belongs to the case of ill-posed problem, in which artificial oscillations are easily arisen in the retrieved energy spectrum even by small errors and noises; hence, it should be treated with appropriate manners.

Aside from the detector response, the presence of noise is another obstacle in the data analysis. In noisy data, template-based signal extraction techniques are usually adopted\footnote{For instance, template-based matched filtering has been widely used in gravitational wave data analyses \citep[see, e.g.,][]{2016nure.book.....S}.}; however, it is not appropriate approach to the analysis of CCSN neutrinos. This is attributed to the fact that the internal dynamics of CCSN involves a non-linear interplay between fluid dynamics, weak interactions and neutrino transport; they are inherently stochastic and highly progenitor dependent. Instead, many previous studies have employed analytic formula to retrieve spectra of neutrinos, in which the spectrum is assumed to be fully characterized by a set of parameters. The parameters estimation is compatible with statistical methods and the capability has been well studied in previous works (see e.g., \citet{2002PhLB..547...37B,2002PhLB..542..239M,2008JCAP...12..006M,2017JCAP...11..036G,2014PhRvD..89f3007L,2016PhRvD..94b3006L,2018JCAP...04..040G,2018PhRvD..97b3019N}). However, this approach heavily relies on the analytic formula and potentially discards some important characteristics of neutrino signals. Therefore, new approaches for the retrieval of neutrino energy spectrum without analytic formulae is necessary as an independent approach.

In this paper we propose a novel method to retrieve energy spectra for all flavors of neutrinos at a CCSN source, having in mind the use of data on multiple detectors: HK (SK) and DUNE. In the spectrum reconstruction process, we do not take any statistical approaches (these are still possible options, though) but rather adopt a deterministic way with a singular value decomposition (SVD) technique which does not require analytic formulae. It should be also mentioned that any neutrino oscillation models can be applied in principle, although our demonstrations presented in this paper are limited in the simple neutrino oscillation models. After reviewing detector characteristics in Sec.~\ref{sec:detector}, we describe the essential idea of our method in Sec.~\ref{subsec:basic} and then get into the detail in Sec.~\ref{sec:detepro}. Although we mainly focus on the technical aspect of the proposed method in this paper\footnote{Some scientific discussions regarding the spectrum reconstruction have been presented in our previous paper \citep{2020arXiv200705000N}}, we assess the capability of the method by demonstrating spectrum reconstructions to neutrino signals computed by our recent 3D CCSN simulations \citep{2019MNRAS.490.4622N,2019MNRAS.485.3153B} in Sec.~\ref{sec:demonst}. In the demonstration, the expected observed data (EOD) to the theoretical neutrino signals are estimated by using SNOwGLoBES. We appropriately handle smearing effects by detector response and Poisson noise in our analysis. We retrieve the energy spectra of neutrinos at the CCSN source by applying our method to the EOD, and then compare them to the answer (results of our CCSN simulations). We believe that this paper will be an important reference to address the issue of how we can combine observed data on multiple detectors to retrieve the energy spectra for all flavors of neutrinos at a CCSN source, although there still remains work needed to improve the method.

\section{Detector Characteristics}\label{sec:detector} 
Before we get into details of the method, we first summarize detector characteristics relevant to this study; two channels from HK (SK) and one channel from DUNE. We note that this choice is not definitive but rather an example; indeed these channels may be replaced (or supplemented) by others in different detectors. For instances, data of pES in liquid scintillator detectors would be useful to measure $\nu_x$ spectrum as proposed by previous studies \citep[see, e.g.,][]{2002PhRvD..66c3001B,2011PhRvD..83k3006D}; charged current reactions with ${^{12}{\rm C}}$ in the same type of detectors are sensitive to $\nu_e$ \citep{2014arXiv1412.8425L}; coherent elastic neutrino scatterings in dark matter detectors are sensitive to all flavors of neutrinos, i.e., they may be more useful reaction channel to reconstruct the energy spectrum of heavy leptonic neutrinos if the detector volume is tonne-scale \citep[see, e.g.,][]{2016PhRvD..94j3009L}. We note, however, that it requires further study of what combination of channels at different detectors provides the most precise measurement of energy spectra of neutrinos, which may depend on CCSN models, progenitors, distances to the source, etc. We postpone the detailed study in future work.

We adopt two reaction channels at HK (SK) detector, which are IBD-p,
\begin{eqnarray}
\bar{\nu}_e  + p \rightarrow e^{+} + n,
\label{ibdchart}
\end{eqnarray}
and eES,
\begin{eqnarray}
\nu  + e^{-} \rightarrow \nu  + e^{-}.
\label{ibdchart}
\end{eqnarray}
The former corresponds to the primary interaction channel for CCSN neutrinos. The energy and angular dependences of the reaction are well known and the detection procedure has been well established; we identify the events through the photo-multiplier tubes installed in the wall of the tank, which detect Cherenkov lights emitted from positrons produced by the IBD. The latter is, on the other hand, a subdominant reaction channel but has some useful properties: (1) it is sensitive to all flavors of neutrinos including their anti-particles: (2) the scattered electrons flight in the forward directions, which provides a way to untangle eES and IBD events\footnote{Since angular distributions of all IBD counts are almost isotropic, the excess of Cherenkov counts in the angular distribution would reflect the eES contribution.}. It should be also mentioned that the addition of gadolinium to detectors further enhances the sensitivity to distinguish them \citep[see, e.g.,][]{2014PhRvD..89f3007L}. In this paper, we assume full tagging efficiencies for the two channels. It should be noted that eES counts are detected through Cherenkov lights from scattered electrons; implying that the flavor-independent counts of eES reactions are not resolved; hence, we use the flavor-integrated counts as the observed quantity in this study.

In this study, we assume that the available detector scale in SK and HK is 32.5 ktons \citep{2016APh....81...39A} and 220 ktons \citep{2018arXiv180504163H}, respectively. We note that the latter is a factor of a few smaller than that used in previous studies \citep[see, e.g.,][]{2008JCAP...12..006M,2018JCAP...04..040G,2018JCAP...12..006G,2018PhRvD..97b3019N}, indicating that the detection count is smaller than that estimated in the previous studies. We also remark on other reaction channels relevant to Oxygen. They are also useful to retrieve the neutrino spectra from CCSN, in particular for the high energy neutrinos ($\gtrsim 50$ MeV). Those reactions are, however, though to be subdominant to CCSN neutrinos and the interaction rates have large uncertainties, which may be improved in future\footnote{Note that there have been many efforts to estimate accurate cross sections for neutrino-nucleus interactions from both theoretical and experimental approaches. See \citet{2012RvMP...84.1307F} and references therein for a review.}. Although we need to take into account those reactions in real observations, we omit them in this study just for simplicity.

In DUNE \citep{2016arXiv160105471A,2016arXiv160807853A,2020arXiv200806647A}, the detector volume is assumed to be $40$ kton and we employ the charged-current reaction channel with Argon (CCAre)
\begin{eqnarray}
\nu_e  + {^{40}{\rm Ar}} \rightarrow e^{-} + {^{40}{\rm K}^{*}},
\label{ibdchart}
\end{eqnarray}
which provides the most accurate estimation of $\nu_e$ among current- and future-planed detectors. By virtue of the high sensitivity to $\nu_e$, DUNE has been expected to play a pivotal role to probe the neutrino mass hierarchy by observing CCSN neutrinos at the earlier phase of the burst ($\lesssim 20 {\rm ms}$), in which energy-luminosity of $\nu_e$ exceeds $10^{53} {\rm erg/s}$. Meanwhile, the luminosity of other species of neutrinos are more than an order of magnitude smaller than that of $\nu_e$, indicating that the detection count at DUNE would be very sensitive to the neutrino oscillation model and the mass hierarchy. For instance, in the case of the normal-mass hierarchy with taking into account adiabatic Mikheyev–Smirnov–Wolfenstein (MSW) effects, the sharp rise of $\nu_e$ event rate on CCAre would disappear due to the substantial mixing of $\nu_x$ \citep[see e.g.][]{2020arXiv200705000N}. In this paper, on the other hand, we suggest to utilize the advantage of DUNE in another way. It is used to distinguish flavor-dependent counts on the eES events at HK (SK), which enable us to retrieve the energy spectrum for all flavor of neutrinos (including $\nu_x$) at a CCSN source. More details will be given in Sec.~\ref{sec:method}.

We assume that background noises can be neglected for the two detectors, whereas Poisson noise is taken into account in this study. This seems to be a reasonable assumption for HK (SK) and DUNE to the analysis of neutrinos from a CCSN \citep[see, e.g.,][for more details]{2007ApJ...669..519I,2019PhRvC..99e5810Z}, although the impact of background noise may need to be considered for more quantitative arguments \citep[see, e.g.,][]{2016ApJ...830L..11A,2016arXiv160607538S,2019ApJ...881..139S}.

\section{Spectrum reconstruction}\label{sec:method} 
\subsection{Basic idea}\label{subsec:basic} 
Here we outline the core of our method how energy spectra of all flavors of neutrinos at a CCSN source can be retrieved by using observed data of HK (SK) and DUNE. In this study, we assume that all heavy leptonic neutrinos and their anti-particles have an identical spectrum at a CCSN source\footnote{Although it is a reasonable approximation to CCSN neutrinos, this is not true in reality. This is attributed to the fact that neutrino-matter interactions are not exactly the same among them; for instance, the correction of weak magnetism is different between neutrinos and anti-neutrinos \citep[see e.g.][]{2002PhRvD..65d3001H} and the difference increases with the energy of neutrinos. Analyses at high energy neutrinos require different methods from those presented in this paper, which are being currently undertaken (Nagakura and Hotokezaka in prep).}. Under the assumption, neutrino fluxes at the Earth can be expressed with taking into account flavor conversions as \citep[see also][]{2000PhRvD..62c3007D}
\begin{eqnarray}
&F^{\rm i}_{e} (\varepsilon) = p^{\rm i} (\varepsilon) F^{0}_{e} (\varepsilon) + \left(1- p^{\rm i} (\varepsilon) \right) F^{0}_{x} (\varepsilon),& \label{eq:flavconv_nue} \\
&\bar{F}^{\rm i}_{e} (\varepsilon) = \bar{p}^{\rm i} (\varepsilon) \bar{F}^{0}_{e} (\varepsilon) + \left(1- \bar{p}^{\rm i} (\varepsilon) \right) \bar{F}^{0}_{x} (\varepsilon),& \label{eq:flavconv_nueb} \\
&F^{\rm i}_{x} (\varepsilon) = \frac{1}{2}  \left(1- p^{\rm i} (\varepsilon) \right) F^{0}_{e} (\varepsilon)
+ \frac{1}{2} \left(1+ p^{\rm i} (\varepsilon) \right) F^{0}_{x} (\varepsilon),& \label{eq:flavconv_nux} \\
&\bar{F}^{\rm i}_{x} (\varepsilon) = \frac{1}{2}  \left(1- \bar{p}^{\rm i} (\varepsilon) \right) \bar{F}^{0}_{e} (\varepsilon)
+ \frac{1}{2}  \left(1+ \bar{p}^{\rm i} (\varepsilon) \right) \bar{F}^{0}_{x} (\varepsilon),& \label{eq:flavconv_nuxb}
\end{eqnarray}
where $F$, $p$ and $\varepsilon$ denote the number flux (fluence), survival probability and energy of neutrinos, respectively; $F^{0}$ corresponds to the neutrino flux without flavor conversions; "$\bar{{\rm A}}$" represents $\rm{A}$ of anti-neutrinos; the subscripts indicates neutrino species; the superscript "${\rm i}$" distinguishes detectors\footnote{The survival probability of neutrinos are different among detectors since Earth matter effects depend on the position of detector at Earth \citep{2001NuPhB.616..307L}.}, i.e., it is either HK (SK) or DUNE in this study. We note that $F^{\rm i}_{x}$ and $\bar{F}^{\rm i}_{x}$ are distinguished in our method, although many previous works treat them collectively. As we shall discuss below, the distinction is necessary, since they are not identical at the Earth\footnote{This can be understood by Eqs.~\ref{eq:flavconv_nux}~and~\ref{eq:flavconv_nuxb}. $F^{0}_e$ and $\bar{F}^{0}_e$ ($\nu_e$ and $\bar{\nu}_e$ at the CCSN source, respectively) are, in general, different from each other, implying that $F^{\rm i}_{x}$ and $\bar{F}^{\rm i}_{x}$ are also different.} and their cross section to eES is also different from each other.

The goal of our method is to obtain the energy spectrum of $F^{0}$ for all flavor of neutrinos and then we convert them to those at a CCSN source\footnote{We assume that the distance to a CCSN is known.}. Although the detailed procedure is complex (see Sec.~\ref{sec:detepro}), the concept is quite straightforward. Given the survival probability of neutrinos ($p$) and that of anti-neutrinos ($\bar{p}$), i.e., assuming a neutrino oscillation model, there remain three unknown variables in Eqs.~\ref{eq:flavconv_nue}--~\ref{eq:flavconv_nuxb}; $F^{0}_{e}$, $\bar{F}^{0}_{e}$ and $F^{0}_{x} (= \bar{F}^{0}_{x})$. This implies that three independent observed data with different flavor sensitivity is required to solve the equations. As the first step, we retrieve the energy spectra of $\nu_e$ at DUNE and $\bar{\nu}_e$ at HK (SK) independently from the data of CCAre and IBD-p channels, respectively, by employing a singular value decomposition (SVD) technique (see Sec.~\ref{sec:unfold} for more details). They provide $F^{{\rm DUNE}}_{e}$ and $\bar{F}^{{\rm HK (SK)}}_{e}$, which corresponds to the left hand side of Eq.~\ref{eq:flavconv_nue} with $i={\rm DUNE}$ and that of Eq.~\ref{eq:flavconv_nueb} with $i={\rm HK (SK)}$, respectively. \footnote{Note that those two equations are degenerated only if $p^{\rm{DUNE}} = \bar{p}^{\rm{HK (SK)}}=0$. However, it is not realistic; hence, we do not consider the peculiar case in this paper.}

To close the equations, we need one more data, which is given by the data of eES channel. We introduce a trial flux to one of $F^{0}$s, and then we iteratively search the solution which the flux satisfies the data of eES channel (within a certain error). During the iterative process, the energy spectrum of either $\nu_x$ or $\bar{\nu}_x$ at HK (SK) is retrieved by applying SVD unfolding technique to its eES event. We note that the eES event with $\nu_x$ or $\bar{\nu}_x$ can be estimated by subtracting other species contribution from the total (see Sec.~\ref{sec:detepro} for more details)\footnote{The idea of subtraction technique to determine flavor-dependent eES counts has been already proposed in the literature \citep[see, e.g., ][]{2018PhRvD..97b3019N}}. It should be also emphasized that $\nu_e$ has a dominant contribution in the eES counts, since it interacts electrons through not only neutral current reactions but also those of charged current. As a result, the cross section of eES with $\nu_e$ is several times higher than that with $\nu_x$. This indicates that the small error in the energy spectrum of $\nu_e$ substantially affects the estimation of $\nu_x$ or $\bar{\nu}_x$ eES events; hence the accurate spectrum retrieval of $\nu_e$ by DUNE is highly valuable for this purpose.

 It should be noted, however, that the obtained eES events with $\nu_x$ and $\bar{\nu}_x$ by the subtraction process would be inevitably noisy, even if the accurate subtraction can be made. The Poisson noise of the total eES counts is $\sim \sqrt{N}$, where N denotes the species-integrated counts of eES channel, meanwhile the eES counts with $\nu_x$ (or $\bar{\nu}_x$) is $\sim 1/6$ of the total one, implying that its signal-to-noise ratio (SN-r) is $\sim \sqrt{N}/6$. In a rough estimation, the SN-r of energy-integrated eES counts with $\nu_x$ in HK is less than $10$ for a CCSN at a distance of $10$ kpc. Note also that the energy-dependent SN-r is further reduced with roughly a factor of $\sqrt{n_{\varepsilon}}$ (where $n_{\varepsilon}$ denotes the number of energy grid points), implying that our method may be only applicable to nearby CCSNe. For such a noisy data, statistical methods may be powerful approach, for instance, \citet{2018PhRvD..97b3019N} proposed a spectrum retrieval method by chi-squared statistical analysis with employing analytic formula of neutrino spectrum. This method would be useful and may be compatible with ours, although we postpone further studies in future work.

\subsection{Details of the procedure}\label{sec:detepro} 
In this section, we describe a detailed procedure of our method. We start with retrieving $\nu_e$ and $\bar{\nu}_e$ spectra by applying the SVD unfolding technique to the data of CCAre on DUNE and IBD-p on HK (SK), respectively; the procedure is straightforward, and then we obtain $F^{{\rm DUNE}}_{e}$ and $\bar{F}^{{\rm HK (SK)}}_{e}$ consequently. As the next step, we proceed the retrieval of energy spectrum of $\nu_x$ or $\bar{\nu}_x$, which is carried out with a fixed-point iteration method. As the preparation, we first determine which $F^{0}$ is selected as a trial neutrino flux. Although the final solution does not depend on the selection, it affects the convergence of the iteration. It is determined by comparing four quantities expressed with survival probability of neutrinos: $p^{{\rm DUNE}}$, ($1-p^{{\rm DUNE}}$), $\bar{p}^{{\rm HK (SK)}}$ and ($1-\bar{p}^{{\rm HK (SK)}}$)\footnote{If we consider the case with energy-dependent survival probabilities, we suggest to take an average over the energy.}. Below, we first describe the detail of the selection procedure and the reason why those four quantities are relevant to the convergence of the iteration.

If $p^{{\rm DUNE}}$ is the smallest among the above four quantities (this corresponds to a neutrino oscillation model of adiabatic MSW effect in the case with normal-mass hierarchy), we select $F^{0}_e$ as the trial variable; the reason is as follows. The energy spectrum of $\nu_e$ retrieved from the data of CCAre at DUNE ($F^{{\rm DUNE}}_{e}$) gives a relation between those of $\nu_e$ ($F^{0}_e$) and $\nu_x$ ($F^{0}_x$) at the CCSN source (see Eq.~\ref{eq:flavconv_nue}). If $p^{{\rm DUNE}}$ is very small, $F^{0}_{x}$ (=$\bar{F}^{0}_{x}$) is almost identical to $F^{{\rm DUNE}}_{e}$. By using the $\bar{F}^{0}_{x}$, we can also retrieve $\bar{F}^{0}_e$ from Eq.~\ref{eq:flavconv_nueb} with $\bar{F}^{{\rm HK (SK)}}_{e}$ which is the energy spectrum of $\bar{\nu}_e$ retrieved from the data of IBD-p at HK (SK). These facts indicate that $F^{0}_{x}$ (=$\bar{F}^{0}_{x}$) and $\bar{F}^{0}_e$ are mostly associated with the data of CCAre at DUNE and IBD-p at HK (SK), and $F^{0}_e$ should be determined mainly through the data of eES channel; hence, $F^{0}_e$ is selected as the trial variable.

 In the case that $\bar{p}^{{\rm HK (SK)}}$ is the smallest (this corresponds to a neutrino oscillation model of adiabatic MSW effect in the case with inverted-mass hierarchy), $\bar{F}^{0}_e$ is appropriate to be selected as a trial variable. In this case, $\bar{F}^{{\rm HK (SK)}}_{e}$ provides a close solution of $\bar{F}^{0}_{x}$ ($= F^{0}_{x}$) but being not sensitive to $\bar{F}^{0}_e$ due to the small $\bar{p}^{{\rm HK (SK)}}$ (see Eq.~\ref{eq:flavconv_nueb}). By using $F^{0}_{x}$, $F^{0}_e$ can be also determined from Eq.~\ref{eq:flavconv_nue} with $F^{{\rm DUNE}}_{e}$. This indicates that $F^{0}_{x}$ (=$\bar{F}^{0}_{x}$) and $F^{0}_e$ are well constraint from the data of CCAre and IBD-p channels, and $\bar{F}^{0}_e$ is determined mainly by data of eES channel; hence we select $\bar{F}^{0}_e$ as a trial variable. Finally, if either ($1-p^{{\rm DUNE}}$) or ($1-\bar{p}^{{\rm HK (SK)}}$) is the smallest, we select $F^{0}_x(=\bar{F}^{0}_x)$ as a trial variable.

We can start the iterative process by setting the trial flux to be zero initially. We confirm that all cases in our demonstrations achieve convergence of the iteration by virtue of the appropriate selection of the trial variable. As the next step, we compute the rest of $F^{0}$s through Eqs.~\ref{eq:flavconv_nue}~and~\ref{eq:flavconv_nueb} by using $F^{{\rm DUNE}}_{e}$ and $\bar{F}^{{\rm HK (SK)}}_{e}$. Here, we emphasize an important fact that the energy spectra of all flavor of neutrinos at Earth can be uniquely determined, albeit tentative, from the three $F^{0}$s. The determination of $\bar{F}^{{\rm HK (SK)}}$ of all flavor of neutrinos enable us to estimate the flavor-dependent eES events, although the flavor-integrated eES channels computed from the $\bar{F}^{{\rm HK (SK)}}$s is not consistent with data of eES channel; hence we iteratively search the trial flux of $F^{0}$ to be consistent with it. The detailed iterative procedure depends on the neutrino oscillation model, meanwhile the essence is basically common among all the cases. Thus, we first provide the procedure in the case that $p^{{\rm DUNE}}$ is the smallest, and then we briefly mention in other cases with focusing on the difference.

In the case that $p^{{\rm DUNE}}$ is the smallest, $F^{{\rm DUNE}}_{e}$ and $\bar{F}^{{\rm HK (SK)}}_{e}$ provide close solutions of $F^{0}_{x}$, $\bar{F}^{0}_{x}$ and $\bar{F}^{0}_{e}$; indeed, they are exact if $p^{{\rm DUNE}}$ is zero. This indicates that $\bar{F}^{{\rm HK (SK)}}_{x}$ computed from Eq.~\ref{eq:flavconv_nuxb} is also close to the actual energy spectrum of $\bar{\nu}_x$ at HK (SK), i.e., it is not necessary to use spectrum unfolding techniques to obtain the spectrum. Similarly, we can estimate $F^{{\rm HK (SK)}}_{e}$ by an algebraic relation of Eq.~\ref{eq:flavconv_nue} with $F^{0}_{e}$ and $F^{0}_{x}$. Note that $F^{{\rm HK (SK)}}_{e}$ is close to $F^{{\rm DUNE}}_{e}$, unless the Earth matter effect significantly affects neutrino flavor conversions.

By using the retrieved energy spectra of $\nu_e$, $\bar{\nu}_e$ and $\bar{\nu}_x$ at HK (SK), we estimate a number of eES counts with each flavor of neutrinos. In this study we use SNOwGLoBES, in which the energy spectrum of those neutrinos is injected to the eES channel of HK (SK), which provides the flavor-dependent eES counts as a function of energy. We subtract the sum of those eES counts from the flavor-integrated one, which provides the eES counts with $\nu_x$. We then apply the SVD method to obtain $F^{{\rm HK (SK)}}_{x}$. By using the $F^{{\rm HK (SK)}}_{x}$ and $F^{0}_{x}$, we compute $F^{0}_{e}$ from Eq.~\ref{eq:flavconv_nux} with $i={\rm HK (SK)}$. The obtained $F^{0}_{e}$ is a solution if it is the same as the one we set in a priori as a trial flux.

We adopt a fixed-point iteration method, in which the trial flux ($F^{0}_{e}$) is updated with the one obtained through the above procedure. Despite the fact that the method does not guarantee the convergence of the iteration, it works quite well, which is by virtue of the appropriate selection of the trial variable. To see it more clearly, let us consider the case that $p^{{\rm DUNE}}$ is zero. In this case, $F^{0}_{x}$, $\bar{F}^{0}_{x}$ and $\bar{F}^{0}_{e}$ can be computed from $F^{{\rm DUNE}}_{e}$ and $\bar{F}^{{\rm HK (SK)}}_{e}$ (see Eq.~\ref{eq:flavconv_nue}); meanwhile $F^{0}_{e}$ is nothing to do with those observed data. Instead, the observed data of eES with $\nu_x$ (see Eq.~\ref{eq:flavconv_nux}) is fully responsible for the determination of $F^{0}_{e}$. If we further assume that the Earth matter effect in flavor conversion is neglected, we can obtain $F^{0}_{e}$ without iteration. This fact indicates that, in the case with finite value of $p^{{\rm DUNE}}$ (but small), the obtained $F^{0}_{e}$ by the proposed procedure is close to the solution, i.e., the convergence would be achieved easily\footnote{We note that this is a multi-variables root-finding problem (the number of unknown variables corresponds to that of the energy mesh in the spectrum of neutrinos), indicating that it is, in general, not easy to achieve the convergence.}. In the case of the normal-mass hierarchy with adiabatic MSW effects, we find that the convergence is achieved with a few iterations.

Below we describe the procedure in other cases with different survival probabilities. As mentioned already, it is essentially the same as that described above; hence we only focus on the difference from the above case. In the case that $\bar{p}^{{\rm SK/IBD}}$ is the smallest, $\bar{F}^{0}_{e}$ is selected as the trial value. During the iteration, we use the SVD to eES counts with $\bar{\nu}_x$ to obtain the energy spectrum of $\bar{\nu}_x$ and search the consistent solution of $\bar{F}^{0}_{e}$ by using a fixed-point iteration method. For the case that either ($1-p^{{\rm DUNE}}$) or ($1-\bar{p}^{{\rm HK (SK)}}$) is the smallest, on the other hand, we need to iteratively search a solution of $F^{0}_{x} (= \bar{F}^{0}_{x})$. In this case, however, there are no preferable choices which we carry out a spectral inversion by SVD to eES counts by $\nu_x$ or $\bar{\nu}_x$; thus we can pick up one of them.

Here we need to mention a caveat. In Sec.~\ref{sec:demonst}, we demonstrate the capability of our method, meanwhile they are only for simple neutrino oscillation models, which are adiabatic MSW effect in the case with normal- or inverted-mass hierarchy. This indicates that our demonstration is limited only in the case that either $p^{{\rm DUNE}}$ or $\bar{p}^{{\rm HK (SK)}}$ is the smallest; hence we do not know whether the fixed-point iteration method works in other neutrino oscillation models. Thus, it may be necessary to use more stable multi-variable root-finding algorithms \citep[see, e.g.,][]{2018arXiv180904495O}; the issue will be addressed in future work.

\subsection{Unfolding procedure}\label{sec:unfold} 
We adopt a singular value decomposition (SVD) technique to retrieve energy spectrum of neutrinos from observed data of each reaction channel. The SVD unfolding technique has been widely used in various fields, and the procedure has been well described in the literature. Hence, we refer readers to previous studies for the detail \citep[see, e.g.,][]{1996NIMPA.372..469H,2019PhRvD..99l3009L}. Instead, we describe other essential parts of the procedure in this section; for instance, we present a new adaptive energy-gridding technique, which is compatible with the SVD technique. We note that there still remains work needed to improve the technique, which will be done in collaboration with data analysts of each detector in future.

The unfolding process can be formulated with a matrix equation,
\begin{eqnarray}
A \hspace{0.5mm} {\bf x} = {\bf b},
\label{eq:resmatdef}
\end{eqnarray}
where $A$, ${\bf x}$ and ${\bf b}$ denote the response matrix, energy distribution (spectrum) of neutrinos and measurements, respectively\footnote{We note that SNOwGLoBES provides angular-integrated counts in each detector. In real data analyses, the angular distribution will be available for some reaction channels, which will provide useful information on the signal.}. Note that ${\bf x}$ is directly associated with the number flux (fluence) of neutrino at each detector (see in Eqs.~\ref{eq:flavconv_nue}--~\ref{eq:flavconv_nuxb}). In our demonstration (see Sec.~\ref{sec:demonst}), we retrieve time-integrated energy spectrum of neutrinos\footnote{To study the time dependent signal, the time integration is carried out with a certain time bin.}, i.e., 
\begin{eqnarray}
x_i = \int F(t,\varepsilon_i) \hspace{0.5mm} dt,
\label{eq:xidef}
\end{eqnarray}
where $\varepsilon_i$ denotes the grid-center energy of $i$-th bin.

We prepare a common energy grid for ${\bf x}$ and ${\bf b}$ vectors; i.e., the number of elements (dimensions) in the both vectors is $n_{\varepsilon}$. This also implies that $A$ is a matrix with $n_{\varepsilon} \times n_{\varepsilon}$. In this study we set a uniform energy grid of $n_{\varepsilon}=100$ zones to cover from $0$ MeV to $100$ MeV. Hereafter, we refer to the energy grid with a "base grid". We note that the base grid is not directly used to the SVD computations but rather be used as a reference to construct the adaptive grids (see below).

We start with generating response matrices of all reaction channels by using SNOwGLoBES. In the software, smearing effects by detector responses are included and the energy resolution of each detector is also taken into account in the output. The response matrices can be constructed by computing detection counts to each reaction channel for the following neutrino energy spectrum,
\begin{eqnarray}
&&x_{j} = 1, \nonumber \\
&& \hspace{-6mm}  x_{k} = 0 \hspace{2mm} {\rm for ~all~ } k(\neq j).
\label{eq:singleeneinjec}
\end{eqnarray}
The output, energy-dependent counts on each channel, is stored into ${\bf b}$ and it gives $a_{ij}$ as
\begin{eqnarray}
a_{ij} = b_i.
\label{eq:resmatdef}
\end{eqnarray}
By running from $j=1$ to $j=n_{\varepsilon}$, we obtain all the elements of the matrix A.

According to the instruction of SNOwGLoBES, however, the data format of input/output in SNOwGLoBES is fixed; for instance, the input data has a format with 501 rows for neutrino energy ranging from $0.1$ MeV to $100.1$ MeV uniformly (i.e., the cell width is $0.2$ MeV), and it is 200 rows from $0$ MeV to $100$ MeV (not exactly uniformly but the cell width is roughly $\sim 0.5$ MeV) for the output. This indicates that the energy grid is not the same as those we adopt in our method. Hence, we need to define a rule to remap energy spectra from one grid to another.

Let us consider to remap an energy spectrum from grid A to grid B. We first compute the number of the signal on each energy bin of grid A. In the case that a bin of grid A is completely covered by a bin of grid B, the number of the signal on the bin of grid A is simply added to the bin of grid B. On the other hand, if a bin of grid A is partially overlapped with a bin of grid B, the signal on the bin of grid A is added to the bin of grid B with a weight. The weight is determined with an assumption that the energy spectrum is flat inside the bin, i.e., it is given by the ratio of the energy width of the overlapped region to that of the bin of grid A \citep[see also][for the similar technique]{2014ApJS..214...16N}. This rule uniquely determines remapping an energy spectrum between the two grids, and it also guarantees that the number of energy-integrated signal is exactly conserved.

As discussed in \citet{1996NIMPA.372..469H}, the re-scaling of the matrix equation may improve the accuracy of the spectrum reconstruction. We define the normalization factor by using the matrix elements,
\begin{eqnarray}
c_{i} = \sum_{j=1}^{n_{\varepsilon}} a_{ij}.
\label{eq:normmateq}
\end{eqnarray}
It should be noted, however, that some elements of the vector ${\bf c}$ are zero due to the energy threshold of each reaction channel. This implies that we can not directly use ${\bf c}$ to normalize the matrix equation. As shown below, however, this issue is resolved with our newly-developed adaptive energy-gridding technique.

In real observations, energy-dependent detected signals are input data, and they are expressed with ${\bf b}$ in our formulation. In this study, we estimate ${\bf b}$ by using SNOwGLoBES based on our 3D CCSN neutrino signals. It should be noted, however, that the neutrino signal in our model does not contain noise (since it is purely a theoretical signal), which is not realistic. Hence, we generate Poisson noise to each energy cell\footnote{Since we have already known the noiseless number of events in our theoretical models, the Poisson noise can be easily generated.} and add them into the ${\bf b}$ in our demonstrations (see Sec.~\ref{sec:demonst}). Note also that, for the case of spectrum reconstruction of $\nu_x$ (or $\bar{\nu}_x$) from the eES channel, the Poisson noise is computed from the flavor-integrated eES events. As described in Sec.~\ref{sec:detepro}, we compute ${\bf b}$ of $\nu_x$ (or $\bar{\nu}_x$) by taking the subtraction process from the flavor integrated ${\bf b}$ included the Poisson noise.

\begin{figure}
  \begin{minipage}{1.0\hsize}
        \includegraphics[width=\columnwidth]{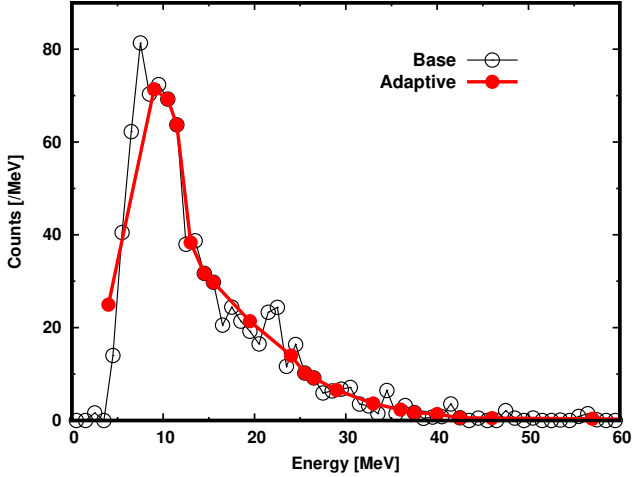}
    \caption{Energy spectrum of time-integrated event counts on the base grid (black) and adaptive (red) one. The corresponding reaction channel is eES with $\nu_x$s. We select a 3D 19 $M_{\sun}$ model in our CCSN simulations and the neutrino signal is computed under the assumption of normal-mass hierarchy with adiabatic MSW effect. See the text for more details.}
    \label{graph_adaptivemesh}
  \end{minipage}
\end{figure}

Although we can compute ${\bf x}$ by applying the SVD method to the matrix A with ${\bf b}$, we exercise an ingenuity in our method. This effort is motivated by a fact that the accuracy of the retrieval of neutrino energy spectrum depends on the energy grid. If the bin width is too small (i.e., high resolution in the energy spectrum), the Poisson noise in each bin tends to be high and may even dominate the signal. In this case, the resultant energy spectrum of neutrinos would be very noisy or yield artifacts due to extensive stabilization treatments in the SVD method, implying that the outcome would be worthless. On the other hand, if the bin width is too large, the coarse energy resolution smears out the energy-dependent feature of the signal. Therefore, it is important to set appropriate energy grids to maximize the accuracy of the spectrum retrieval. We also note that, since the SN-r of neutrino observations varies with various factors such as reaction channels and distance to the CCSN source etc., the energy grid needs to be adapted accordingly.

Hence, we develop an adaptive energy-gridding technique; the procedure is described below. As mentioned already, we have already set the "base grid", which provides the finest energy resolution in the spectrum. We note that our procedure guarantees that, if the SN-r is very high, the energy grid becomes identical to the "base grid" except for the bins with low energy and null measurements (see below for more details). At first, we search energy bins with the number of measurement is zero (i.e., $b_i=0$), and then combine the corresponding bin to a neighbor cell of the lower energy side. Next, we further bundle the energy bins in the range of $0$~MeV to $E_{\rm min}$. This treatment stabilizes the SVD method, since the event counts in the low energy region would be smaller than those at the region around $\sim 10$ MeV due to the less neutrino flux and the smaller cross-sections of the reaction rate on each detection channel, which cause large statistical fluctuations and the result in destabilizing the spectrum retrieval. Although the coarse energy resolution at the low energy region discards extracting detailed energy-dependent features in the spectrum, we prioritize the stability here. $E_{\rm min}$ is determined based on the detector response of each reaction channel and we set $E_{\rm min}=6, 7$ and $8$ MeV for IBD-p, CCAre and eES, respectively. Finally, we further combine the energy bins by imposing a condition that the number of counts per energy (NCPE) has a single-peak in the spectrum. We compute the NCPE on each bin from low-energy side, and if the second peak emerges in the spectrum, we combine the corresponding bin to the neighboring cell of the lower energy side. We repeat the procedure until the NCPE in the spectrum has a single-peak.

In Fig.~\ref{graph_adaptivemesh}, we show an energy spectrum of eES event counts with $\nu_x$ (NCPE) by using two different energy grid\footnote{We pick up the spectra from the demonstration which we present in Sec.~\ref{sec:demonst}.}; the black open-circles correspond to the energy spectrum expressed with the base grid, and the red filled-circles are the counterparts with the adaptive energy grid. As shown in the figure, oscillatory features emerge in the spectrum on the entire energy regime for the case with the base grid (which is due to the Poisson noise). On the contrary, those features are smeared out in the case with the adaptive grid. We confirm that this treatment stabilizes the spectral retrieval by the SVD algorithm; indeed, the retrieved energy spectrum of neutrinos becomes smooth consequently.

We make a remark on a single-peak condition in the procedure, since the real observation may have multiple peaks in the spectrum. In this case, energy bins would be combined excessively and it is inevitable to miss catching multiple peaks. We note, however, that the obtained spectrum is still informative to study neutrino signals (since we do not change the raw observed data), and it will be possible to judge whether the multiple peaks in the spectrum are real signal or emerging due to statistical noises. If the peaks are recognized as signals\footnote{This determination hinges on the confidence level of the analysis.}, we can redo the same analysis with decomposing the energy grid at the corresponding region in the spectrum. Although it can be automated in our analysis pipeline, it is much beyond the scope of this paper.

\begin{figure}
  \begin{minipage}{1.0\hsize}
        \includegraphics[width=\columnwidth]{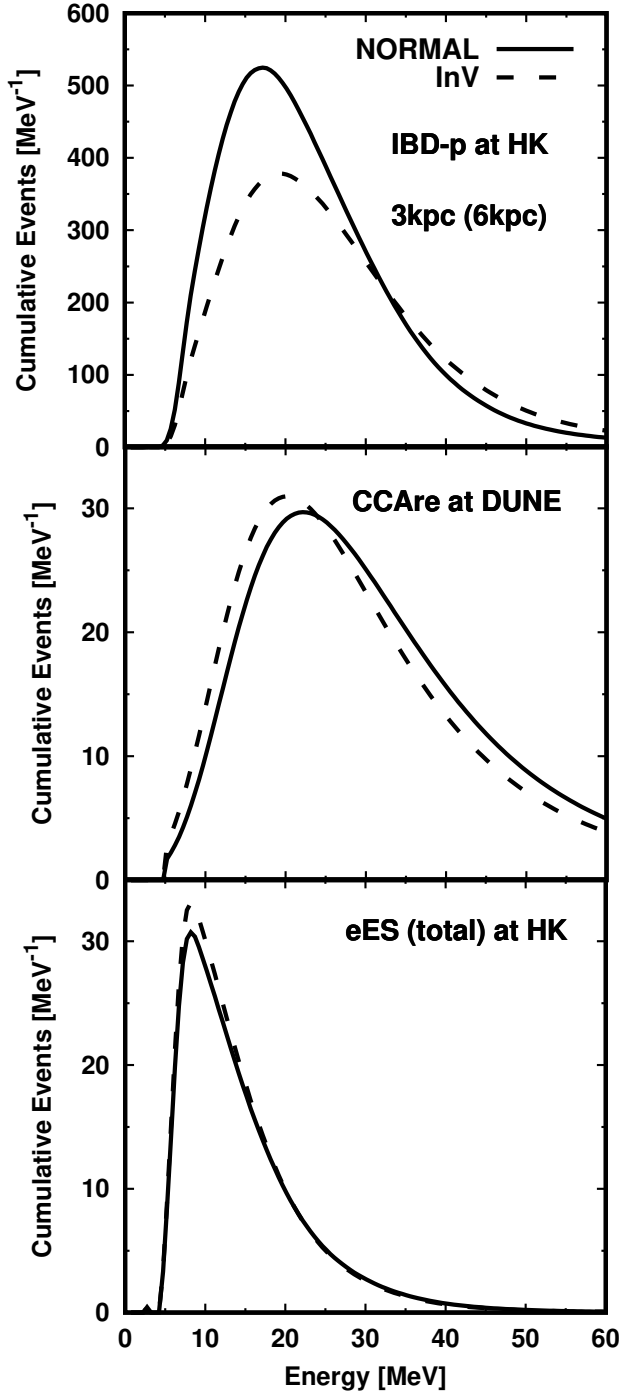}
    \caption{Energy spectrum of event counts in each reaction channel for our CCSN model; from top to bottom IBD-p at HK, CCAre at DUNE and eES (flavor-integrated counts) at HK. The line type represents the neutrino oscillation model. In the spectrum, smearing effects due to detector responses are included, meanwhile Poisson noise will be added later (see text for details). The distance to the CCSN source is assumed to be 3 kpc, indicating that the corrected distance is $6$ kpc.}
    \label{graph_detectionspectrum}
  \end{minipage}
\end{figure}

In accordance with the change of the energy grid, the matrix equation (Eq.~\ref{eq:resmatdef}) needs to be modified. To do this, we define a mapping function $G$, which connects to the index of the bin in the base grid to that for the adaptive one,
\begin{eqnarray}
I = G(i),
\label{eq:mapaddress}
\end{eqnarray}
where $i$ (lowercase) and $I$ (uppercase) denote the index of bin for the base grid and that for the adaptive one, respectively. We then determine a response matrix $S$ for the adaptive grid as
\begin{eqnarray}
s_{IJ} = \sum_{i,j} a_{ij} \hspace{1.0mm} \delta_{G(i),I} \hspace{1.0mm} \delta_{G(j),J} \hspace{1.0mm},
\label{eq:newmateleme}
\end{eqnarray}
where $s_{IJ}$ and $\delta$ denotes the matrix elements of $S$ and the Kronecker delta, respectively.
Similarly, the elements of the source- and normalized vectors can be obtained as
\begin{eqnarray}
d_{I} = \sum_{i} b_{i} \hspace{1.0mm} \delta_{G(i),I} \hspace{1.0mm}, \label{eq:newsourceeleme} \\
h_{I} = \sum_{i} c_{i} \hspace{1.0mm} \delta_{G(i),I} \hspace{1.0mm}.
\label{eq:newsourceeleme}
\end{eqnarray}
Since there are no zero components in ${\bf h}$, we can apply it to normalize the matrix equation.
The normalized matrix $S$ and the vector ${\bf d}$ can be written as
\begin{eqnarray}
&& \hspace{-20mm} \tilde{s}_{IJ} = \frac{s_{IJ}}{h_{I}}, \\
&& \hspace{-20mm} \tilde{d}_{I} = \frac{d_{I}}{h_{I}},
\label{eq:normalizationmatveceleme}
\end{eqnarray}
and the matrix equation can be rewritten as
\begin{eqnarray}
\tilde{S} \hspace{0.5mm} {\bf y} = {\bf \tilde{d}},
\label{eq:matnormalized}
\end{eqnarray}
where ${\bf y}$ represents the energy distribution of neutrinos on the adaptive energy grid. We apply the SVD method to Eq.~\ref{eq:matnormalized} and then obtain ${\bf y}$.

Finally, we make a few remarks. As mentioned already, the adaptive energy grid varies with the reaction channel, which is not a convenient property for joint analysis among different detectors. Hence, we compute ${\bf x}$ of each reaction channel by linearly interpolating the logarithm of ${\bf y}$. Another remark is that we find a large error in the spectrum at the low energy regime ($\lesssim 6$ MeV) regardless of reaction channel, which is mainly due to the low sensitivity of detectors (energy threshold) and large Poisson noise. Since we are not interested in such a low energy region in our method, we take an ad hoc prescription to suppress large errors. In the energy region with $\lesssim 6$ MeV, we modify the spectrum by interpolating linearly from $x$ at 6 MeV. Although this is a crude prescription, we confirm that it does not compromise the accuracy of our analysis, since the typical energy of CCSN neutrinos is higher than 10 MeV.

\section{Demonstration}\label{sec:demonst} 

\begin{figure*}
  \begin{minipage}{0.9\hsize}
        \includegraphics[width=\columnwidth]{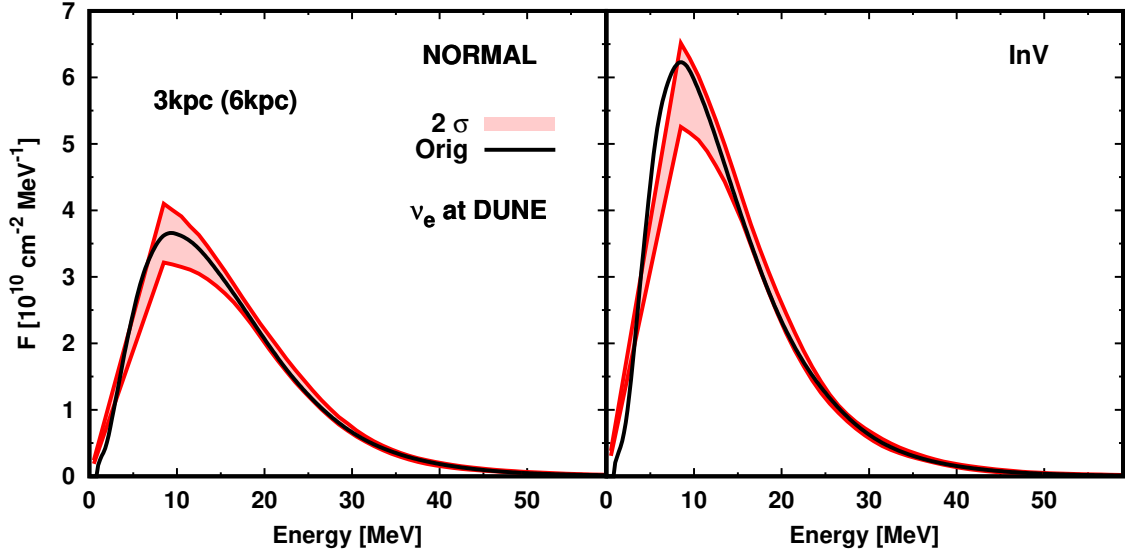}
    \caption{Energy spectrum of $\nu_e$ flux (fluence) at DUNE, which are retrieved by applying the SVD technique to the observed data by CCAre reaction channel. The black line corresponds to the original (true) spectrum, while the red shaded region correspond to the retrieved spectrum with $2 \sigma$ confidence level. The source distance is assumed to be 3 kpc (the corrected one is 6 kpc). The left and right panels show the result with the case of normal- and inverted-mass hierarchy, respectively.}
    \label{graph_nuespectrum_byCCAreDUNE}
  \end{minipage}
\end{figure*}

\subsection{CCSN model}\label{subsec:model} 
We test the capability of our method by demonstrating the retrieval of neutrino energy spectra. We employ theoretical neutrino data of a 19 $M_{\sun}$ 3D CCSN model\footnote{See \citet{2020arXiv200705000N} for the progenitor dependence.} simulated by a neutrino-radiation hydrodynamic code, F{\sc{ornax}} \citep{2019ApJS..241....7S}, in which essential input physics of CCSN is incorporated. The neutrino transport is solved with multi-group two-moment method, in which the effects of fluid-velocity and gravitational redshift in neutrino transport are approximately handled. The details of the fluid dynamics and neutrino emissions are discussed in \citet{2019MNRAS.490.4622N,2019MNRAS.489.2227V,2020MNRAS.491.2715B,2020MNRAS.492.5764N}. In the post bounce evolution, the model goes through a stalled accretion shock phase and then transits into a phase of runaway shock expansion at $\sim 400$ ms after the core bounce. Since there are no artifices in the explosive dynamics, the essential characteristics of fluid dynamics and the neutrino signal of CCSN would be captured adequately.

\begin{figure*}
  \begin{minipage}{0.9\hsize}
        \includegraphics[width=\columnwidth]{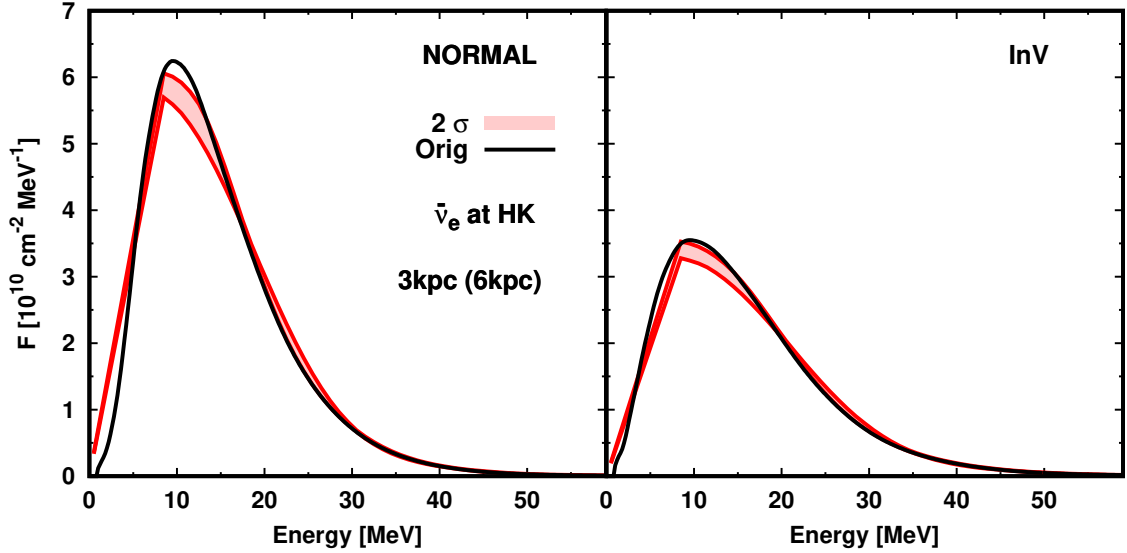}
    \caption{The same as Fig.~\ref{graph_nuespectrum_byCCAreDUNE} but for $\bar{\nu}_e$ retrieved from observed data of IBD-p channel at HK.}
    \label{graph_nuebspectrum_byIBDpHK}
  \end{minipage}
\end{figure*}

\begin{figure*}
  \begin{minipage}{0.9\hsize}
        \includegraphics[width=\columnwidth]{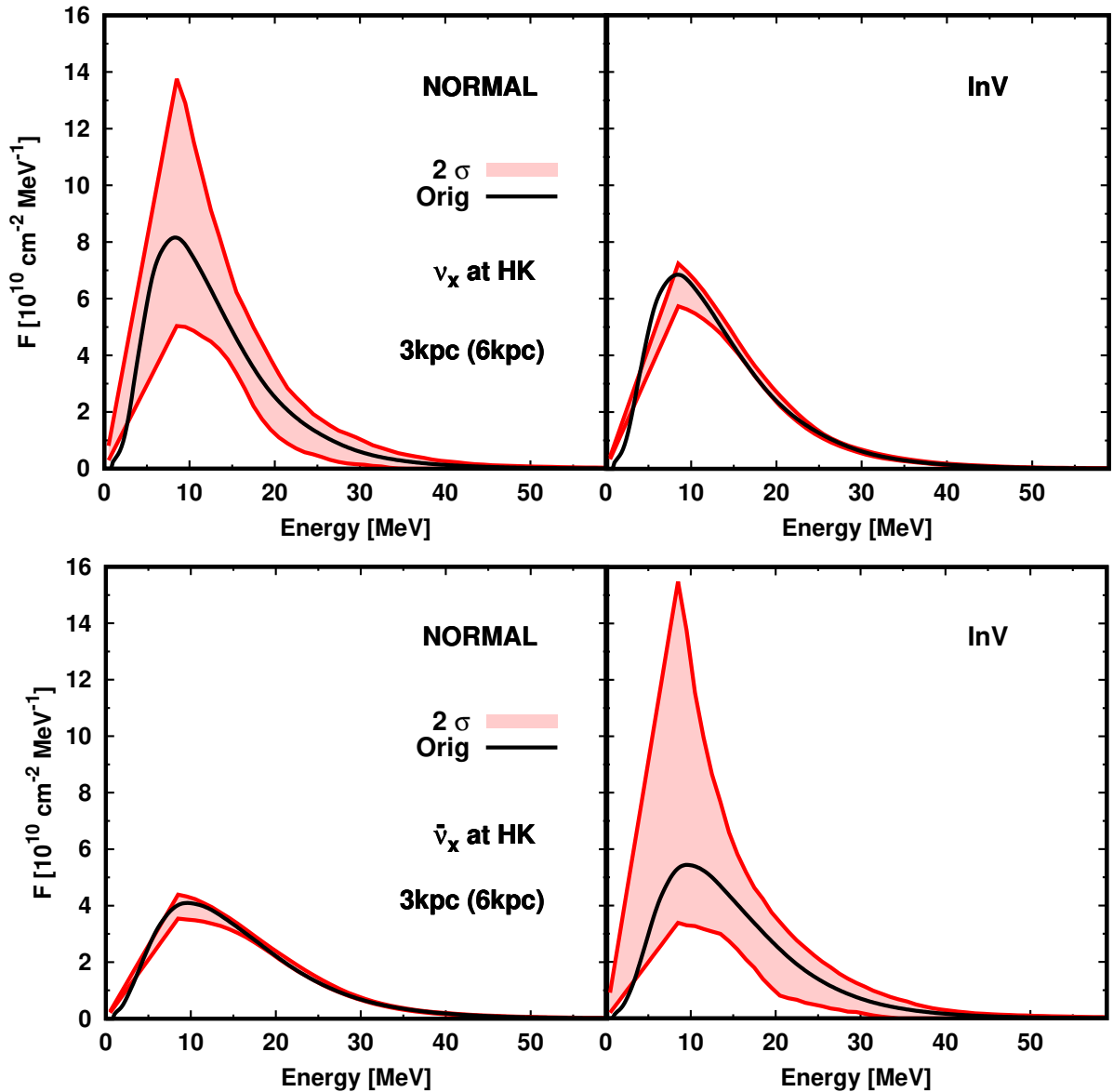}
    \caption{The same as Fig.~\ref{graph_nuebspectrum_byIBDpHK} but for $\nu_x$ (top) and $\bar{\nu}_x$ (bottom) at HK.}
    \label{graph_nux_nuxbspectrum_atHK}
  \end{minipage}
\end{figure*}

We note that the neutrino data of our CCSN models is publicly available\footnote{from the link, \url{https://www.astro.princeton.edu/~burrows/}}. In this demonstration, we use the time-integrated energy spectra of neutrinos at the end of the simulation ($871$ ms after the bounce), i.e., our CCSN model covers only the accretion phase of CCSN but does not include PNS cooling phase. This indicates that the total neutrino energy (TONE) in our model is roughly 4 times smaller than the actual total energy of neutrinos from CCSN. It implies that our method is capable of retrieving energy spectra of neutrinos with the same accuracy at $\sim 2$ times greater distance. In the following demonstrations, the corrected distance will be also displayed as a reference.

In this paper, we focus only on two simplified cases of neutrino oscillation models, which are adiabatic MSW models in normal- and inverted-mass hierarchy. Given the neutrino oscillation model, we compute the energy spectra of event counts in each reaction channel; IBD-p at HK, CCAre at DUNE and eES (flavor-integrated) at HK, which are displayed in Fig.~\ref{graph_detectionspectrum}. The source distance is assumed to be 3 kpc\footnote{We note again that our CCSN model only covers the accretion phase, indicating that including also the cooling could allow similar up to approximately double the distance, i.e., the corrected distance is 6 kpc (displayed with parentheses in the figure).}. The smearing effects due to detector responses are taken into account in the plots. Given the event counts, Poisson noise are computed and then added to the spectrum in each realization. In this study, 1000 Poisson noise realizations are performed, and we discuss the capability of our method with $\sim 2 \sigma$ confidence level. 

\subsection{Retrieving energy spectrum of neutrinos}\label{subsec:demoret}

\begin{figure*}
  \begin{minipage}{0.9\hsize}
        \includegraphics[width=\columnwidth]{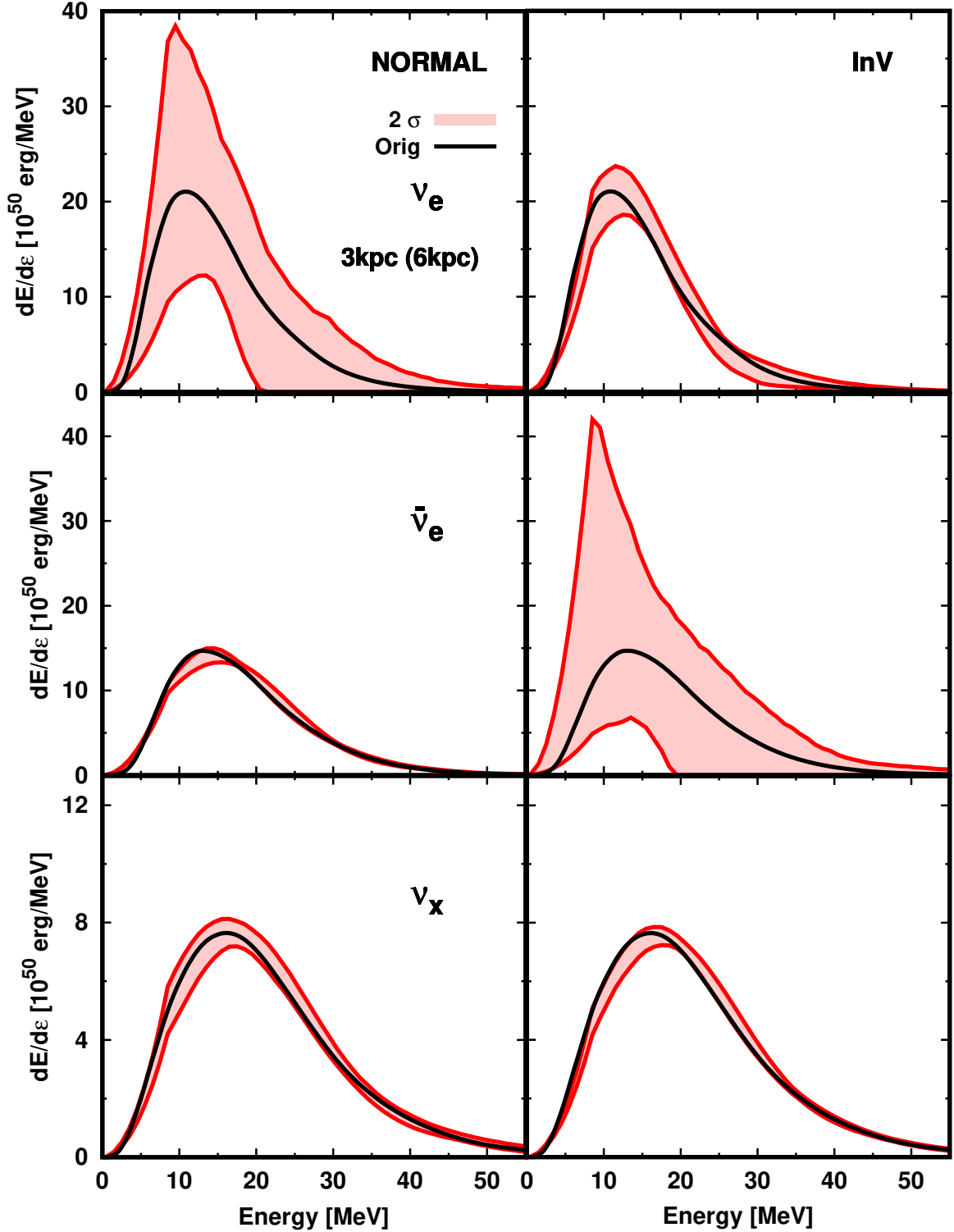}
    \caption{The time-integrated energy spectrum for all flavors of neutrinos at the CCSN source retrieved by our method. From top to bottom, $\nu_e$, $\bar{\nu}_e$, and $\nu_x$, respectively. Left and right panels correspond to the case of normal- and inverted-mass hierarchy.}
    \label{graph_retenespectsource_3kpc}
  \end{minipage}
\end{figure*}

\begin{figure*}
  \begin{minipage}{0.9\hsize}
        \includegraphics[width=\columnwidth]{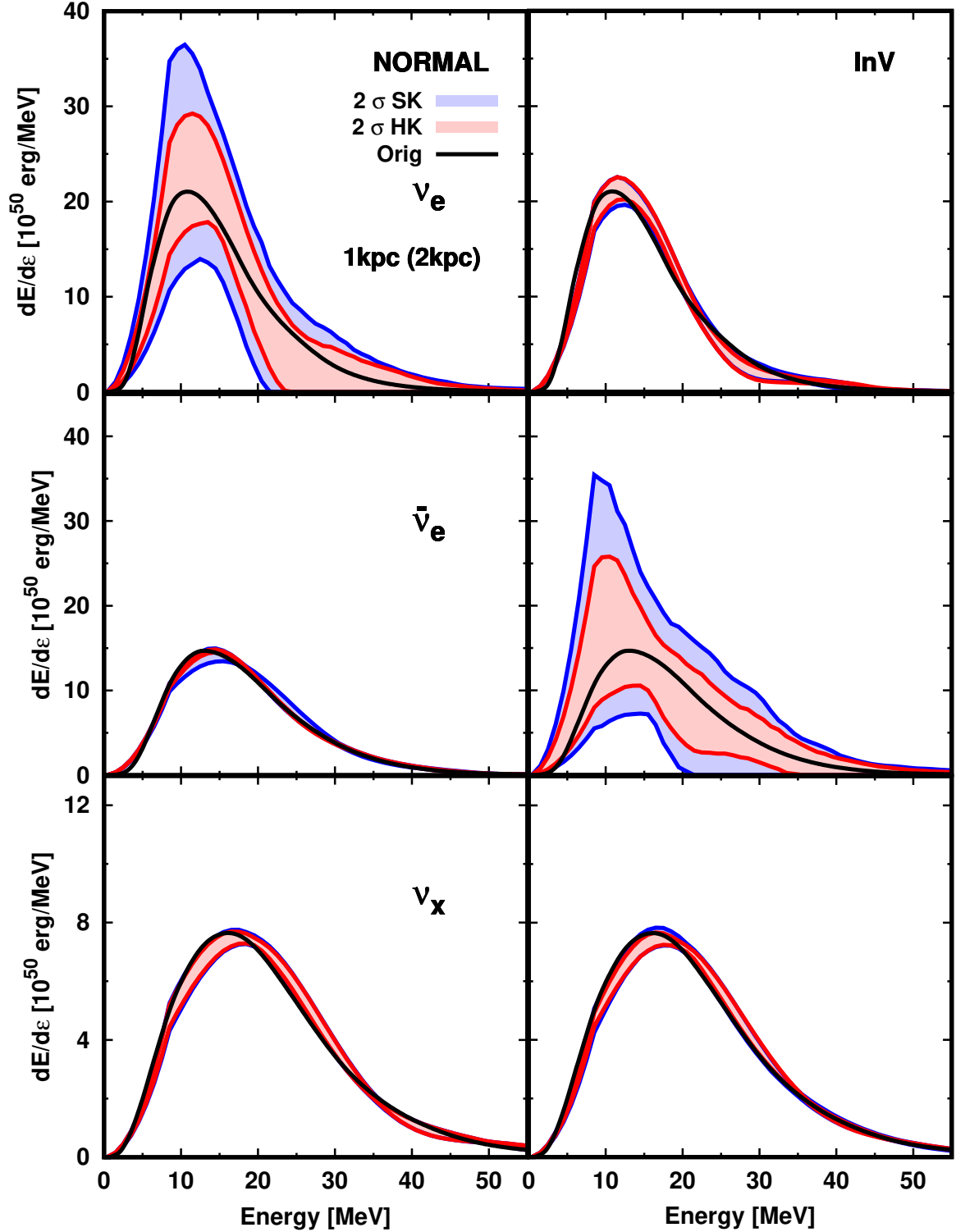}
    \caption{The same as Fig.~\ref{graph_retenespectsource_3kpc} but the distance to the CCSN source is assumed to be 1(2) kpc. For the blue color, we employ expected observed data of IBD-p and eES on SK instead of those on HK.}
    \label{graph_retenespectsource_1kpc}
  \end{minipage}
\end{figure*}

\begin{figure*}
  \begin{minipage}{0.9\hsize}
        \includegraphics[width=\columnwidth]{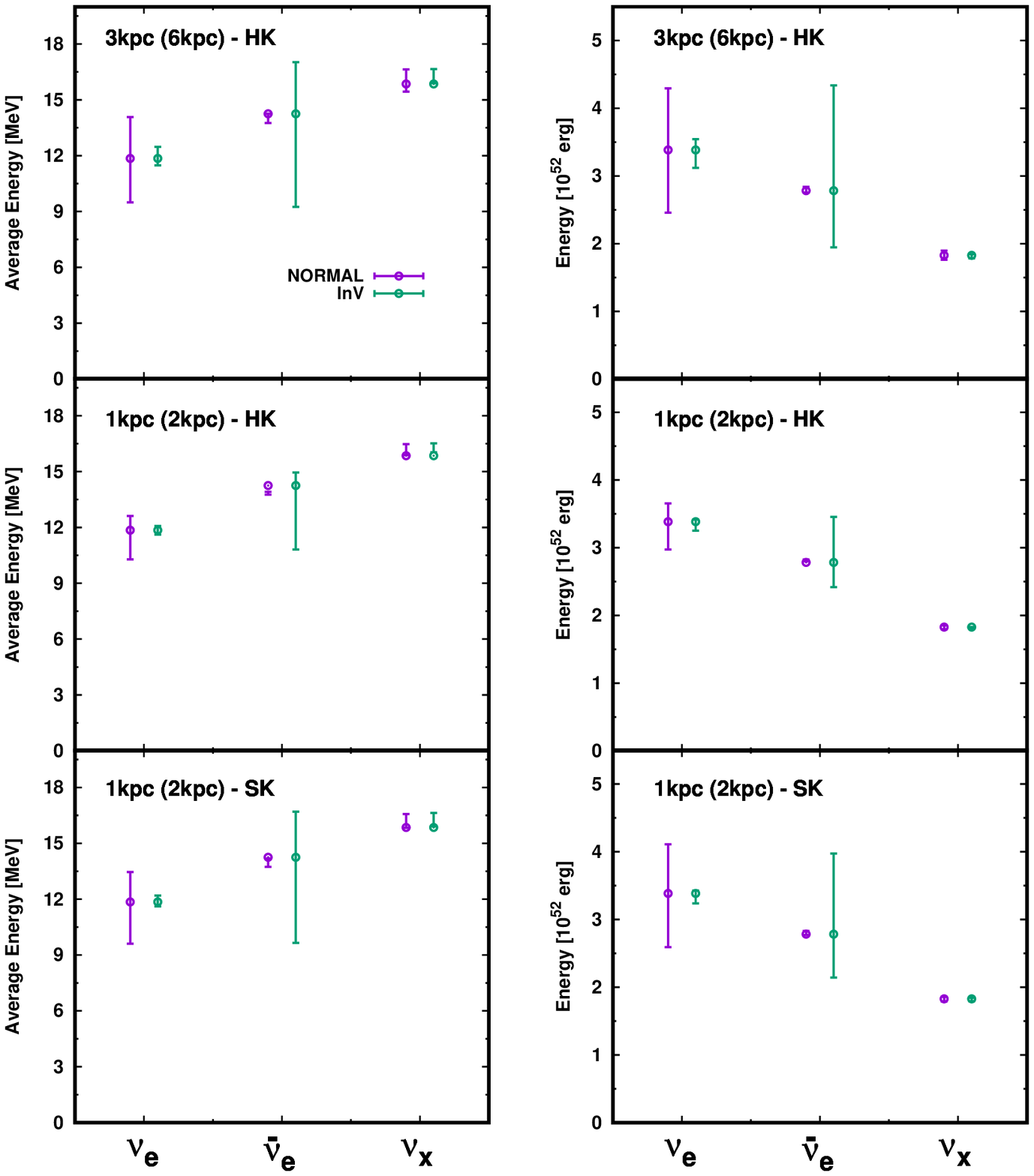}
    \caption{The retrieved average energy (left) and total energy (right) for each flavor of neutrino. Purple and green denote the case of normal- and inverted-mass hierarchy, respectively. The open circles correspond to the answer (i.e., results from our CCSN simulations), and the error bar corresponds to $2 \sigma$ confidence level. In the top panel, we show the result in the case with employing HK and the distance to the source is assumed to be 3(6) kpc. In the middle one, we change the distance to 1(2) kpc. In the bottom, we change the detector configuration of HK to that of SK in the demonstration and the source distance is assumed to be 1(2) kpc.}
    \label{graph_error_Reconst_aveE_Etot_19M}
  \end{minipage}
\end{figure*}

Figures~\ref{graph_nuespectrum_byCCAreDUNE} and \ref{graph_nuebspectrum_byIBDpHK} display retrieved spectrum of $\nu_e$ flux (fluence) at DUNE and that of $\bar{\nu}_e$ at HK, respectively. These are obtained by applying the SVD method (see Sec.~\ref{sec:unfold}) to the data of CCAre at DUNE and IBD-p at HK, respectively. We find that there are relatively large errors around $10$ MeV. This is mainly due to a systematic error of detector response. As a common property among all reaction channels, they are less sensitive to low-energy neutrinos, indicating that small noises in event counts yield large errors in the retrieved spectrum\footnote{As pointed out in Sec.~\ref{sec:detepro}, we, hence, replace the retrieved spectrum at very low energy region ($< 6$ MeV) to that interpolated from the value at $6$ MeV.}. The error is, however, less than $20 \%$ and $10 \%$ for $\nu_e$ and $\bar{\nu}_e$ fluxes, respectively, indicating that the retrieved energy spectra agree reasonably well with the solution of the spectrum.

Figure~\ref{graph_nux_nuxbspectrum_atHK} displays retrieved energy spectra of $\nu_x$ (top) and $\bar{\nu}_x$ (bottom) at HK. As described in Sec.~\ref{sec:detepro}, we apply the SVD method to the data of eES events to retrieve $\nu_x$ ($\bar{\nu}_x$) for the case of normal- (inverted-) mass hierarchy. Since the data of eES with heavy leptonic neutrinos is very noisy, they are retrieved with lower accuracy than those of $\nu_e$ and $\bar{\nu}_e$ (see Sec.~\ref{sec:detepro} for more details). On the other hand, the spectrum of $\bar{\nu}_x$ ($\nu_x$) in the case of normal- (inverted-) mass hierarchy is retrieved from algebraic relations of Eq.~\ref{eq:flavconv_nuxb} (Eq.~\ref{eq:flavconv_nux}), indicating that its precision is mainly determined with the data of CCAre of DUNE and IBD-p of HK; hence, the resultant spectrum has the similar precision with those of $\nu_e$ and $\bar{\nu}_e$.

Fig.~\ref{graph_retenespectsource_3kpc} corresponds to the final goal in our method, i.e., it portrays the retrieved energy spectrum for all flavor of neutrinos at the CCSN source. In the case with normal-mass hierarchy, the survival probability, $p$, is nearly zero, indicating that $\nu_e$ spectrum retrieved from CCAre at DUNE is mainly responsible for high precision $\nu_x$ spectrum at the CCSN source (see Sec.~\ref{subsec:basic} for more details). This also indicates that $\bar{\nu}_x$ spectrum at the source is retrieved precisely. It allows us to reconstruct $\bar{\nu}_e$ at the CCSN source by using the $\bar{\nu}_e$ spectrum retrieved from IBD-p at HK, which is also done with a high precision. The large error found in $\nu_e$ energy spectrum at the CCSN source is directly associated with that of $\nu_x$ at HK (see the top and left panel in Fig.~\ref{graph_nux_nuxbspectrum_atHK}), which is mainly retrieved from the eES channel with $\nu_x$. For the case with inverted-mass hierarchy (right panels), the $\bar{\nu}_e$ spectrum retrieved from IBD-p at HK provides the high precision spectrum for $\bar{\nu}_x$ ($=\nu_x$) at the CCSN source. The $\nu_e$ energy spectrum at the CCSN source is also retrieved precisely by virtue of high precision of $\nu_e$ spectrum at DUNE and $\nu_x$ spectrum at the CCSN source. The large error in $\bar{\nu}_e$ energy spectrum at the CCSN source is mainly due to the low precision of $\bar{\nu}_x$ spectrum retrieved by eES channel with $\bar{\nu}_x$ at HK. These trends which we find in this demonstration is consistent with our expectation described in Sec.~\ref{sec:method}.

As a reference, we make two more demonstrations; one of them is that we change the source distance from 3 kpc to 1 kpc. The results are displayed as red shaded regions in Fig.~\ref{graph_retenespectsource_1kpc}. We find that the error in $\nu_e$ ($\bar{\nu}_e$) energy spectrum for the case of normal- (inverted-) mass hierarchy under the HK detector configuration becomes substantially smaller than those in Fig.~\ref{graph_retenespectsource_3kpc}. This is simply due to the reduction of Poisson noise in eES events at HK by virtue of the shorter distance to the CCSN source. On the other hand, we do not find any substantial improvements to retrieve energy spectra for other species of neutrinos, indicating that systematic errors seem to be dominated for them. However, the error in the spectrum is within a few percents in the energy range of $\lesssim 30$ MeV, which would be sufficient to analyze neutrino signals from CCSN. It should be also noted that there are other ingredients to cause systematic errors; for instance, the uncertainty in the cross-sections of each reaction channel. Indeed, the cross section of CCAre is poorly determined in the low energy neutrinos, and the uncertainty could be one of the major uncertainties in the retrieved spectrum. As such, the further improvements of our method needs to be done with removing these uncertainties. In Fig.~\ref{graph_retenespectsource_1kpc} we display the result of another demonstration; we use a detector configuration of SK instead of that of HK (see blue shaded regions in the figure). We confirm that the spectrum retrieval for all flavors of neutrinos are possible with SK and DUNE, if the CCSN source is very nearby, $\lesssim 1 (2)$ kpc.

Finally, we show the retrieved average energy and total energy for each flavor of neutrino in Fig.~\ref{graph_error_Reconst_aveE_Etot_19M}. The precision of the retrieval to each quantity can be understood through that found in retrieved energy spectrum as discussed above. Importantly, either $\nu_e$ or $\bar{\nu}_e$ has relatively large errors in these quantities, meanwhile those of the rest of the species are determined precisely (even in the case with the SK configuration). This characteristics leads us to an important conclusion that we can estimate the total neutrino energy (TONE) precisely \citep[see also][]{2020arXiv200705000N}. We also find that our method may be capable of investigating the hierarchy of average energy of CCSN neutrinos if the source distance is less than 1(2) kpc and HK detector is available (see the middle and left panel in Fig.~\ref{graph_error_Reconst_aveE_Etot_19M}), although it hinges on neutrino oscillation models.

\section{Summary and Conclusion}\label{sec:sumandconc} 

In this paper, we present a new method of retrieving energy spectra for all flavors of neutrinos by using data with multiple detectors. In our method, we employ data of IBD-p and eES reaction channels at a water Cherenkov detector such as SK and HK, and CCAre of DUNE. At each detector, we retrieve the energy spectrum of neutrinos by using the newly-developed SVD unfolding algorithm with adaptive energy-gridding technique. Given the neutrino flavor conversions, we iteratively search the energy spectra of $\nu_e$, $\bar{\nu}_e$, and $\nu_x$ at the CCSN source, which provides consistent spectrum of event counts on each reaction channel. Although it belongs multi-variables root-finding problem, the fixed-point iteration method works quite well by virtue of the appropriate selection of a trial variable. During the iteration, we need $\nu_x$ and $\bar{\nu}_x$ eES events at SK or HK, which is possible by subtracting contributions of other neutrinos from the total eES events. The $\nu_e$ spectrum retrieved at DUNE plays an important role to make the subtraction precisely, indicating that the joint analysis with data of multiple detectors is highly valuable. In Sec.~\ref{sec:demonst}, we show the capability of our proposed method by applying it to one of theoretical models of CCSN neutrinos provided by our recent 3D CCSN simulations. The retrieved energy spectrum of all flavors of neutrinos agrees reasonably well with a solution of each spectrum, lending confidence to our method. We also note that our method is capable of yielding useful measurements of the spectra as a function of time, although the demonstration is postponed to future work. The time axis information will be helpful to extract physical information on CCSN.

On the other hand, there is a relatively large error in the energy spectrum of either $\nu_e$ ($\bar{\nu}_e$) at the source in the case of normal- (inverted-) mass hierarchy, which is directly associated with large Poisson noise of eES events with $\nu_x$ or $\bar{\nu}_x$. This is apparently the major source of errors in our method, which should be improved in future. One feasible way would be to use other reaction channels having sensitive to heavy leptonic neutrinos; for instances, coherent elastic neutrino scatterings in tonne-scale dark matter detectors potentially improve the precision of spectrum retrieval \citep{2016PhRvD..94j3009L}; the channels of neutral-current interactions are also another candidate \citep[see, e.g.,][]{2002PhRvD..66c3001B,2019PhRvD..99l3009L}, although it will improve only in the energy range above $\sim 20$ MeV, indicating that the relatively large error at $\sim 10$ MeV may not be substantially improved. We currently study how we can effectively combine the data of more than 3 independent reaction channels to improve the precision of the spectrum retrieval.

Last but not least, it is important to test the capability of our method under more complicated neutrino oscillation models. As pointed out in \citet{2019PhRvD.100d3004A,2019ApJ...886..139N,2020PhRvR...2a2046M,2020PhRvD.101b3018D,2020PhRvD.101f3001G,2020PhRvD.101d3016A}, the fast pair-wise neutrino flavor conversion, one of the collective neutrino oscillation models, would commonly occur in the post-bounce phase of CCSNe. We also note that the Earth matter effect should be incorporated in real observations if we analyze data of multiple detectors \citep[see, e.g.,][]{2001NuPhB.616..307L}. These tests are currently underway and will be published elsewhere.

\section*{Acknowledgements}
We acknowledge Adam Burrows, Kate Scholberg, Shunsaku Horiuchi, Kenta Hotokezaka, 
David Vartanyan, and David Radice for useful comments and discussions.
We acknowledge support from the U.S. Department of Energy Office of Science and the Office
of Advanced Scientific Computing Research via the Scientific Discovery
through Advanced Computing (SciDAC4) program and Grant DE-SC0018297
(subaward 00009650). In addition, we gratefully acknowledge support
from the U.S. NSF under Grants AST-1714267 and PHY-1804048 (the latter
via the Max-Planck/Princeton Center (MPPC) for Plasma Physics). 
An award of computer time was provided
by the INCITE program. That research used resources of the
Argonne Leadership Computing Facility, which is a DOE Office of Science
User Facility supported under Contract DE-AC02-06CH11357.
In addition, this overall research
project is part of the Blue Waters sustained-petascale computing project,
which is supported by the National Science Foundation (awards OCI-0725070
and ACI-1238993) and the state of Illinois. Blue Waters is a joint effort
of the University of Illinois at Urbana-Champaign and its National Center
for Supercomputing Applications. This general project is also part of
the ``Three-Dimensional Simulations of Core-Collapse Supernovae" PRAC
allocation support by the National Science Foundation (under award \#OAC-1809073).
Moreover, access under the local award \#TG-AST170045
to the resource Stampede2 in the Extreme Science and Engineering Discovery
Environment (XSEDE), which is supported by National Science Foundation grant
number ACI-1548562, was crucial to the completion of this work.  Finally,
the authors employed computational resources provided by the TIGRESS high
performance computer center at Princeton University, which is jointly
supported by the Princeton Institute for Computational Science and
Engineering (PICSciE) and the Princeton University Office of Information
Technology, and acknowledge our continuing allocation at the National
Energy Research Scientific Computing Center (NERSC), which is
supported by the Office of Science of the US Department of Energy
(DOE) under contract DE-AC03-76SF00098.

\section*{DATA AVAILABILITY}
The data underlying this article will be shared on reasonable request to the corresponding author.

\bibliographystyle{mnras}
\bibliography{bibfile}

\bsp	
\label{lastpage}
\end{document}